\documentclass[twocolumn,english,showpacs,superscriptaddress]{revtex4-1}
\usepackage[T1]{fontenc}
\usepackage[latin9]{inputenc}
\usepackage{geometry}
\usepackage{xcolor}
\usepackage{babel}
\usepackage{units}
\usepackage{braket}
\usepackage{amsmath}
\usepackage{amssymb}
\usepackage{stmaryrd}
\usepackage{graphicx}
\usepackage{wasysym}
\usepackage{bbold}
\usepackage[bookmarks=false]{hyperref}

\makeatletter

\IfFileExists{lmodern.sty}{\usepackage{lmodern}}{}
\usepackage{babel}

\usepackage{hyperref}

\makeatother

\begin{document}
\title{Demonstration of three- and four-body interactions between trapped-ion spins}
\date{\today}

\author{Or Katz}
\email{Corresponding authors: or.katz@duke.edu ; lei.feng@duke.edu}
\address{Duke Quantum Center, Duke University, Durham, NC 27701}
\address{Department of Electrical and Computer Engineering, Duke University, Durham, NC 27708}
\address{Department of Physics, Duke University, Durham, NC 27708}

\author{Lei Feng}
\address{Duke Quantum Center, Duke University, Durham, NC 27701}
\address{Department of Electrical and Computer Engineering, Duke University, Durham, NC 27708}
\address{Department of Physics, Duke University, Durham, NC 27708}

\author{Andrew Risinger}
\address{Department of Electrical and Computer Engineering, University of Maryland, College Park, MD 20742}

\author{Christopher Monroe}
\address{Duke Quantum Center, Duke University, Durham, NC 27701}
\address{Department of Electrical and Computer Engineering, Duke University, Durham, NC 27708}
\address{Department of Physics, Duke University, Durham, NC 27708}
\address{IonQ, Inc., College Park, MD  20740}

\author{Marko Cetina}
\address{Duke Quantum Center, Duke University, Durham, NC 27701}
\address{Department of Electrical and Computer Engineering, Duke University, Durham, NC 27708}
\address{Department of Physics, Duke University, Durham, NC 27708}

\begin{abstract}
Quantum processors use the native interactions between effective spins to simulate Hamiltonians or execute quantum gates. In most processors, the native interactions are pairwise, limiting the 
efficiency of controlling entanglement between many qubits.
Here we experimentally demonstrate a new class of native interactions between trapped-ion qubits, extending conventional pairwise interactions to higher order. 
We realize three- and four-body spin interactions as examples, showing that high-order spin polynomials may serve as a new toolbox for quantum information applications.  
\end{abstract}
\maketitle

\begin{figure*}[t]
\begin{centering}
\includegraphics[width=18cm]{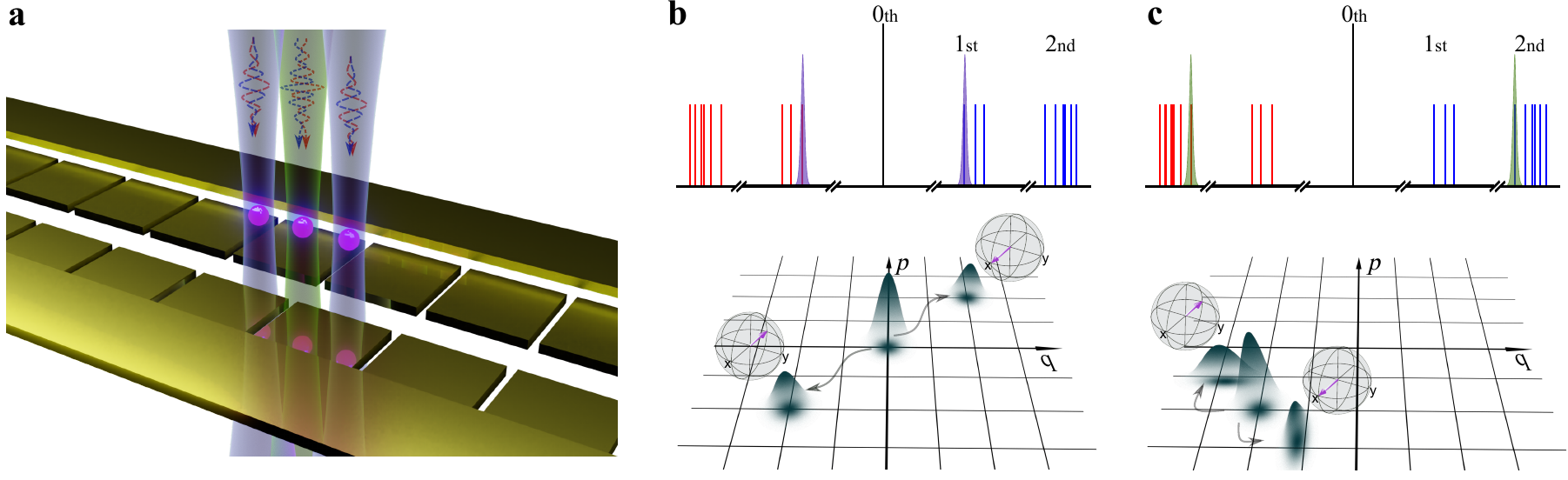}
\par\end{centering}
\centering{}\caption{\textbf{Native operations of a trapped-ion quantum processor.}  \textbf{a}, Linear chain of laser-cooled ion spins in a chip trap. Array of optical beams addressing individual ions enable exquisite control over the state of each spin and its coupling to collective phonon-modes using Raman transitions (additional global beam forming the Raman pairs not shown). \textbf{b}, Spin-dependent displacement of one phonon-mode and its representation over the acoustic phase-space. Simultaneous driving of the first red and blue sideband transitions of one ion near resonance with one phonon mode displaces the phonon wavepacket (grey Gaussian) into two distinct trajectories, depending on the state of that spin along the x direction over its Bloch sphere (purple arrow). \textbf{c}, Spin-dependent squeezing of one phonon-mode. Simultaneous resonant driving of the second red and blue sideband transitions of one ion and one phonon mode squeezes (anti-squeezes) the phonon wavepacket along one direction of phase-space depending on the state of that spin along the x direction. Thin-lines in (\textbf{b}-\textbf{c}) represent the sideband spectrum and narrow red and blue Gaussians represent the spectral content of the Raman coupling near the sidebands. 
\label{fig:exp_overview}}
\end{figure*}

Useful quantum computers and simulators rely upon the controllable generation of quantum entanglement between their elementary constituents, such as qubits or effective spins. Such entanglement allows the efficient exploration of a large state space, which can speed up the computation of certain problems \cite{Mike_and_Ike} or the simulation of the dynamics or phases of model physical systems \cite{Monroe2021}. The generation of entanglement relies on the native interactions between subsets of spins, which in most quantum platforms is pairwise \cite{Molmer1999,kjaergaard2020superconducting,gross2017quantum}. However, higher-order interactions are often featured in Hamiltonian models in nuclear and  high-energy physics  \cite{banuls2020simulating,ciavarella2021trailhead,hauke2013quantum,farrell2022preparations} and spin systems \cite{pachos2004three,muller2011simulating,motrunich2005variational,andrade2022engineering,bermudez2009competing}, as well as quantum circuits and algorithms in quantum chemistry \cite{seeley2012bravyi,o2016scalable,nam2020ground,aspuru2005simulated,hempel2018quantum}, error correction codes \cite{paetznick2013universal,kitaev2003fault}, and other applications \cite{vedral1996quantum, grover1996fast,Wang2001,PhysRevLett.102.040501,Espinoza2021,figgatt2017complete,marvian2022restrictions}. Sequential or parallel application of universal one- and two-body gate sets can, in principle, generate any unitary mapping in Hilbert space that is equivalent to the evolution under high order interactions \cite{Mike_and_Ike}. Yet such constructions carry a large overhead in the number of Trotterization steps \cite{Lloyd1996} or entangling operations \cite{bullock2004asymptotically}, thereby limiting the practical performance of such an approach in the presence of decoherence and noise.

From a fundamental view, few-body ($>$2) interactions can lead to qualitatively different behavior compared to pairwise interactions, as seen across different fields of physics \cite{efimov1970energy,kraemer2006evidence,aaij2015observation}. The study and search for $N$-body interactions has thus become a central research avenue in most quantum platforms, from neutral atoms \cite{weimer2010rydberg,isenhower2011multibit,Levine2019,xing2021realization,khazali2020fast,naidon2017efimov} and superconducting systems \cite{Zahedinejad2015,khazali2020fast,hill2021realization,kim2022high} to chains of trapped atomic ions \cite{CiracZoller1995,monz2009realization,goto2004multiqubit,Espinoza2021,Wang2001}. Yet for trapped-ion processors, which feature dense and controllable qubit connectivity through phonon modes and very long qubit coherence times, robust tunable and scalable interactions beyond the pairwise limit have never been demonstrated.   

Here we demonstrate a new class of native higher-order interactions between qubits in a trapped-ion quantum processor. To do this, we apply a state-dependent squeezing optical drive, which is a simple extension over the conventional state-dependent displacement used for M\o{}lmer-S\o{}rensen (MS) pairwise gates \cite{Molmer1999,Milburn2000,Solano1999}. Such squeezing forces have been applied to trapped-ion systems to improve the performance of pairwise gates \cite{burd2021quantum, shapira2022robust}. Here we instead exploit state-dependent squeezing to generate three- and four-body interactions \cite{Katz2022Nbody, katz2022programmable}. 
We outline avenues to extend the scheme and highlight its potential use for quantum computation and simulation at larger scales.  

The quantum processor, operated at the Duke Quantum Center, is based on a chain of $^{171}\textrm{Yb}^+$ atomic ions confined in a linear Paul trap on a chip \cite{maunz2016high,egan2021fault,Cetina2022}, as shown in Fig.~\ref{fig:exp_overview}a. Each ion represents a qubit or effective spin comprised of two ``clock" levels in its electronic ground-state ($\ket{\uparrow_z}\equiv \ket{F=0,M=0}$ and $\ket{\downarrow_z}\equiv\ket{F=1,M=0}$). We drive motion-sensitive optical Raman transitions on the spin levels using pairs of noncopropagating beams far-detuned from any electronic transitions with a beat-note near the qubit frequency splitting \cite{Olmschenk2007}. 
The spins are initialized and measured using resonant optical pumping and state-dependent fluorescence techniques, resulting in a state preparation and measurement (SPAM) errors of $<0.5\%$ per ion \cite{egan2021scaling}.

The native entangling operations between spins are mediated by phonons and driven by Raman transitions. The phonons reside in collective modes of motion which feature nonlocal and dense connectivity with the spins. We simultaneously drive red and blue sideband transitions to displace or squeeze the motional state of the ions in selective modes \cite{Katz2022Nbody}. Driving the first sideband transitions of the $n$-th ion near the resonance of a single phonon mode generates a spin-dependent displacement $\sigma_{x}^{(n)}\alpha$ where $\sigma_{x}^{(n)}$ is a transverse Pauli matrix and $\alpha$ is the complex displacement parameter. The phonon wave-packet of that mode, represented in the phase-space of its harmonic motion \cite{sorensen2000entanglement,katz2022programmable}, is therefore displaced by $+\alpha$ if the spin points upwards along the $x$ basis but by $-\alpha$ if the spin points downwards, as illustrated in Fig.~\ref{fig:exp_overview}b.

Alternatively, driving the second sideband transitions of the $n$-th ion at twice the resonance frequency of a single phonon mode squeezes the phonon wavepacket by a factor $e^{\xi\sigma_x^{(n)}}$ along the horizontal phonon coordinate $q$ where $\xi$ is given in Eq.~(\ref{eq:xi}); It is squeezed by a factor $e^{-\xi}$ if the spin points downwards along the $x$ axis, but anti-squeezed by $e^{\xi}$ if the spin points upwards along $x$, as shown in Fig.~\ref{fig:exp_overview}c. Crucially, the displacement and squeezing operations depend on the state of the spins, but are independent of the initial phonon state in the Lamb-Dicke regime (when the radial motion along the optical beam is much smaller than the wavelength of the optical drive). Temporal control over the amplitudes and phases of the Raman beams over time $t$ enables full control over $\alpha(t),\,\xi(t)$, as well as determination of the spin axes that diagonlize the spin-dependent forces (See Methods and \cite{Katz2022Nbody,katz2022programmable}). 

Coupling between different spins is realized by accumulation of spin-dependent geometric phase $\Phi$ that shifts the phase of the quantum state as $\left|\psi\right\rangle\rightarrow e^{-i\Phi}\left|\psi\right\rangle$, similar to the underlying mechanism of the pairwise MS interaction. Here the geometric phase is accumulated by a sequence of alternating displacement and squeezing operations which move the phonon wavepacket in closed loops in phase-space. 

\begin{figure*}[t]
\begin{centering}
\includegraphics[width=17.5cm]{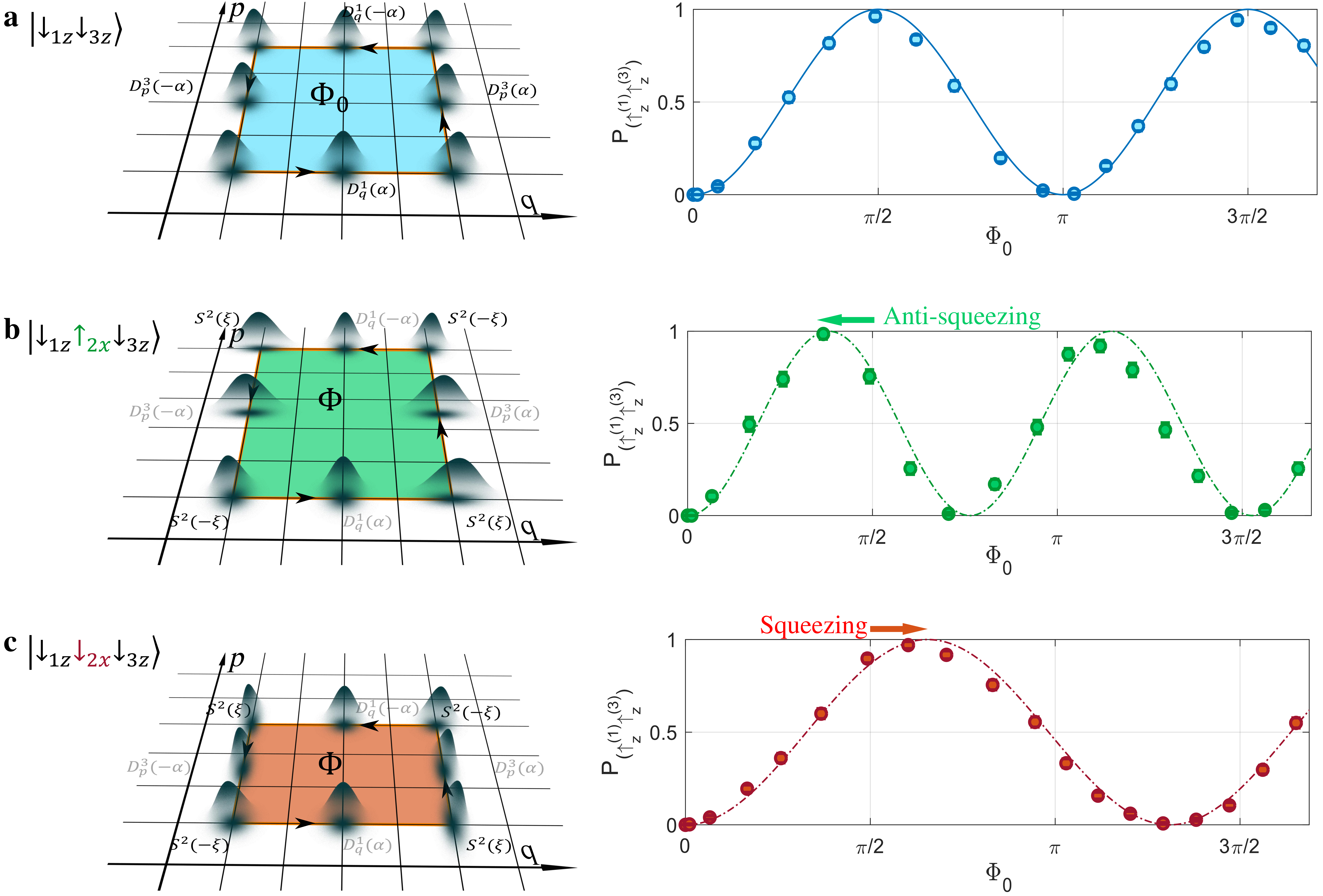}
\par\end{centering}
\centering{}\caption{\textbf{Quantum phase-gates.} \textbf{a}, M\o{}lmer-S\o{}rensen phase-gate between ions number 1 and 3 using displacement operations of one phonon mode using Milburn's scheme \cite{Milburn2000} as illustrated in Fig.~\ref{fig:exp_overview}a. The phase-space area of the enclosed rectangular contour $\Phi_0$ controls the spin evolution, jointly flipping the initial state $\ket{\downarrow_z^{(1)}\downarrow_z^{(3)}}$ into the state $\ket{\uparrow_z^{(1)}\uparrow_z^{(3)}}$. $D^{(n)}_q(\alpha)$ and $D^{(n)}_p(\alpha)$ denote position and momentum displacement operations. \textbf{b-c}, Interspersing spin-dependent squeezing operations $S^{(2)}(\pm\xi)$ on ion number 2 in between displacement stages along the $p$ coordinate scales the accumulated phase-space area $\Phi$ conditioned on the state of that spin (c.f.~Eq.~\ref{eq:Phi_3spins}). (a-c) The phonon wavepacket is brought back to its original state at the end of the gate operation to erase the spin-phonon correlations developed during the gate. Circles represent measured data, bars represent $1\sigma$ binomial uncertainties, and dash-dots lines are the analytically calculated Unitary evolution for the system parameters estimated independently. The applied experimental sequences are presented in \ref{fig:sequences}a-b. \label{fig:phase_space_area}}
\end{figure*}

To demonstrate higher order interactions, we first consider the conventional MS interaction between two ions in a chain of three. Following cooling and spin initialization in the $\ket{\downarrow_z^{(1)}\downarrow_z^{(3)}}$ state via optical pumping, we resonantly drive the lowest-frequency radial phonon mode (``zig-zag'' mode) with a sequence of displacement operations that alternately act on the two edge ions (see \ref{fig:sequences}a), generating a rectangular-shaped loop in motional phase-space as shown in Fig.~\ref{fig:phase_space_area}a \cite{Milburn2000}. The accumulated geometric phase corresponds to the phase-space area enclosed in the loop, which is given by $\Phi=\Phi_0\sigma_{x}^{(1)}\sigma_{x}^{(3)}$. We control $\Phi_0=\alpha^2$ by scaling the amplitude of the displacement pulses while fixing the total duration of the sequence to about $180\,\mu$s. We suppress the displacement of other phonon modes by pulse-shaping of the displacement waveforms \cite{zhu2006trapped} and also suppress the effect of uncompensated level shifts using a pair of echo pulses (see Methods).  Application of this phase-gate jointly flips the spins pair into the state $\ket{\uparrow_z^{(1)}\uparrow_z^{(3)}}$ with probability $p_{(\uparrow_z^{(1)}\uparrow_z^{(3)})}=\sin^{2}{(\Phi_0)}$, which is detected via state-dependent fluorescence and displayed in Fig.~\ref{fig:phase_space_area}a. We determine the scale of $\Phi_0$ by fitting the data in Fig.~\ref{fig:phase_space_area}a to a sine function as a function of the Raman beam intensity. 

We extend the pairwise interaction by interspersing in the sequence squeezing operations that act only on the middle spin and drive predominantly the zig-zag phonon mode (see \ref{fig:sequences}b). These operations are realized as pairs of squeezing and anti-squeezing pulses sandwiching the displacement operations (\ref{fig:phase_space_area}). The squeezing forces scale the momentum displacements by the spin-dependent factor $e^{\xi\sigma_x^{(2)}}\equiv\cosh{(\xi)}\mathbb{1}+\sinh{(\xi)}\sigma_{x}^{(2)}$ where $\mathbb{1}$ is the identity matrix, while nulling the net deformation of the phonon wavepacket. The geometric phase is then given by the scaled rectangular area \begin{equation}\label{eq:Phi_3spins}\Phi=\Phi_0\bigl(\cosh{(\xi)}\sigma_{x}^{(1)}\sigma_{x}^{(3)}+\sinh{(\xi)}\sigma_{x}^{(1)}\sigma_{x}^{(2)}\sigma_{x}^{(3)}\bigr),\end{equation} manifesting two- and three-body interaction terms. 

\begin{figure*}[t]
\begin{centering}
\includegraphics[width=17cm]{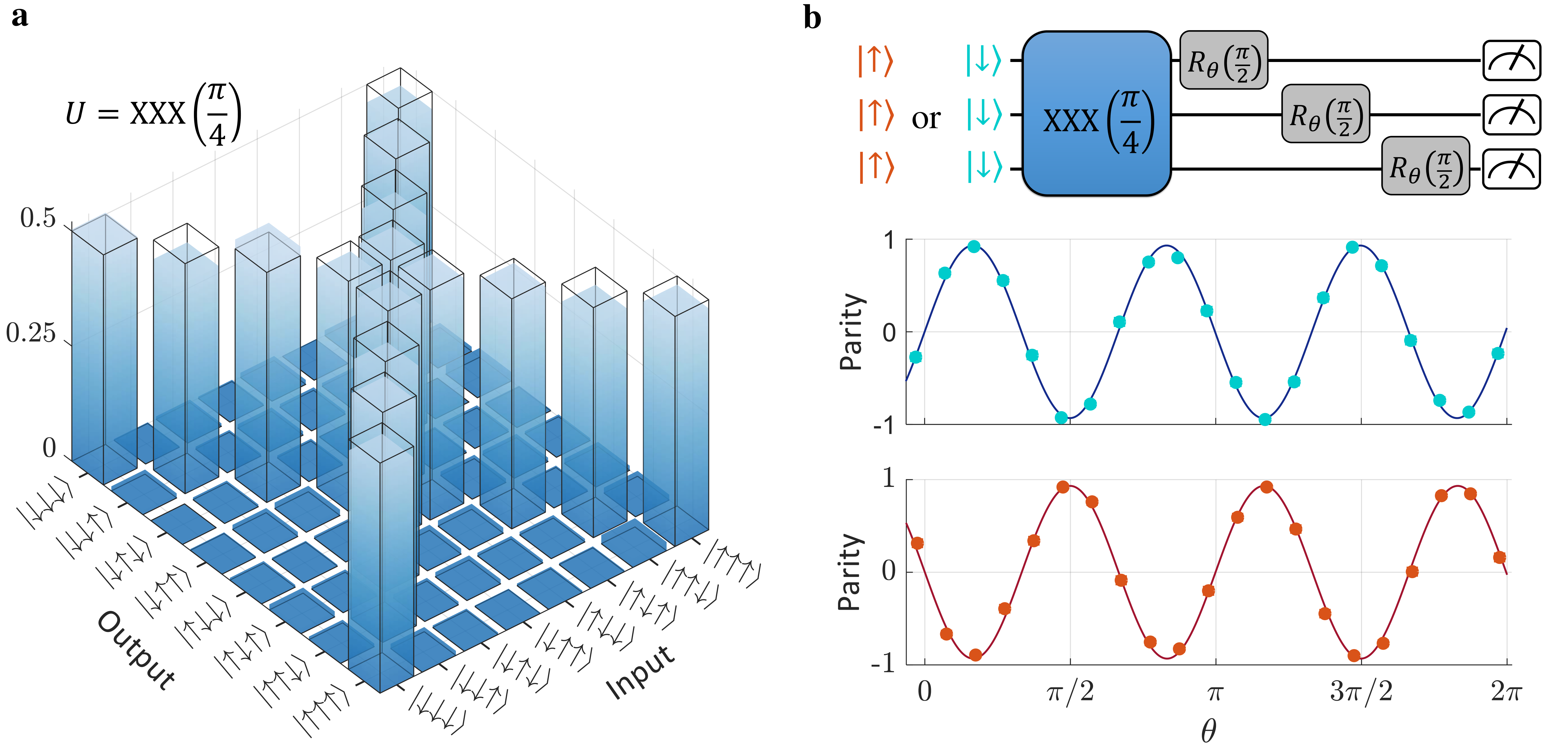}
\par\end{centering}
\centering{}\caption{\textbf{Characterization of a three-body interaction gate.} \textbf{a}, Truth table of a three-body gate ${\textrm{XXX}(\tfrac{\pi}{4})=\exp(-i\tfrac{\pi}{4}\sigma_x^{(1)}\sigma_x^{(2)}\sigma_x^{(3)})}$ gate generated by sequence of displacement and squeezing operations. The input and output spin states are along the z basis. Each input state is ideally mapped into a pair of output states (wire frames), and raw measurement are shown in solid bars. The measured populations of these target states is $(95.8\pm0.9)\%$, averaged over the $8$ measured configurations. \textbf{b}, Characterization of the parity fringe witness observable obtained by the gate. We measure the parity fringe of the output GHZ state for the two initial states $\ket{\downarrow_z^{(1)}\downarrow_z^{(2)}\downarrow_z^{(3)}}$(light blue) and $\ket{\uparrow_z^{(1)}\uparrow_z^{(2)}\uparrow_z^{(3)}}$ (red) with a fitted amplitude of $0.932\pm0.015$ for the two states. The extracted GHZ fidelities for this operation for the two states are $\mathcal{F}=(94.8\pm1.5)\%$ and $\mathcal{F}=(94.4\pm1.9)\%$. Bars represent $1\sigma$ binomial uncertainties and dashed lines are sine functions. The operation $R_{\theta}(\pi/2)$ denotes single qubit rotation by azimuthal angle $\theta$ and a polar angle $\pi/2$ on the Bloch sphere. Measured values are uncorrected for errors in state preparation, measurement, and single qubit rotations.\label{fig:gate_tomo}}
\end{figure*}

We demonstrate the action of this phase-gate in Fig.~\ref{fig:phase_space_area}b-c on the initial states $\ket{\downarrow_z^{(1)}\uparrow_x^{(2)}\downarrow_z^{(3)}}$ and $\ket{\downarrow_z^{(1)}\downarrow_x^{(2)}\downarrow_z^{(3)}}$, for a total sequence time of less than $300\,\mu$s including all displacement, echo and squeezing operations. Similar to the MS interaction, this gate jointly flips the state of the two edge spins, but with probability 
$P_{(\uparrow_z^{(1)}\uparrow_z^{(3)})}=\sin^{2}{(e^{\pm\xi}\Phi_0)}$, whose dependence on $\alpha^2$ is scaled by a factor $e^{\xi}$ ($e^{-\xi}$) and is conditioned on the state of the middle spin pointing upwards (downwards) along the $x$ direction. The calculated evolution of $\xi=0.27$, estimated independently from Eq.~\ref{eq:xi} given the applied optical force amplitude agrees well with observation.

This many-body entanglement operation features full control over the amplitudes of the two- and three-body terms appearing in Eq.~(\ref{eq:Phi_3spins}). We can, for example, eliminate the contribution of the two-body term by setting $\Phi_0=\pi/\cosh{\xi}$ and generate a pure three body term with amplitude $\pi\tanh\xi$. We note that maximally entangled states between three spins in this case requires only $1$ dB of squeezing ($\tanh\xi=\tfrac{1}{4}$) \cite{Katz2022Nbody}. 

\begin{figure*}[t]
\begin{centering}
\includegraphics[width=17cm]{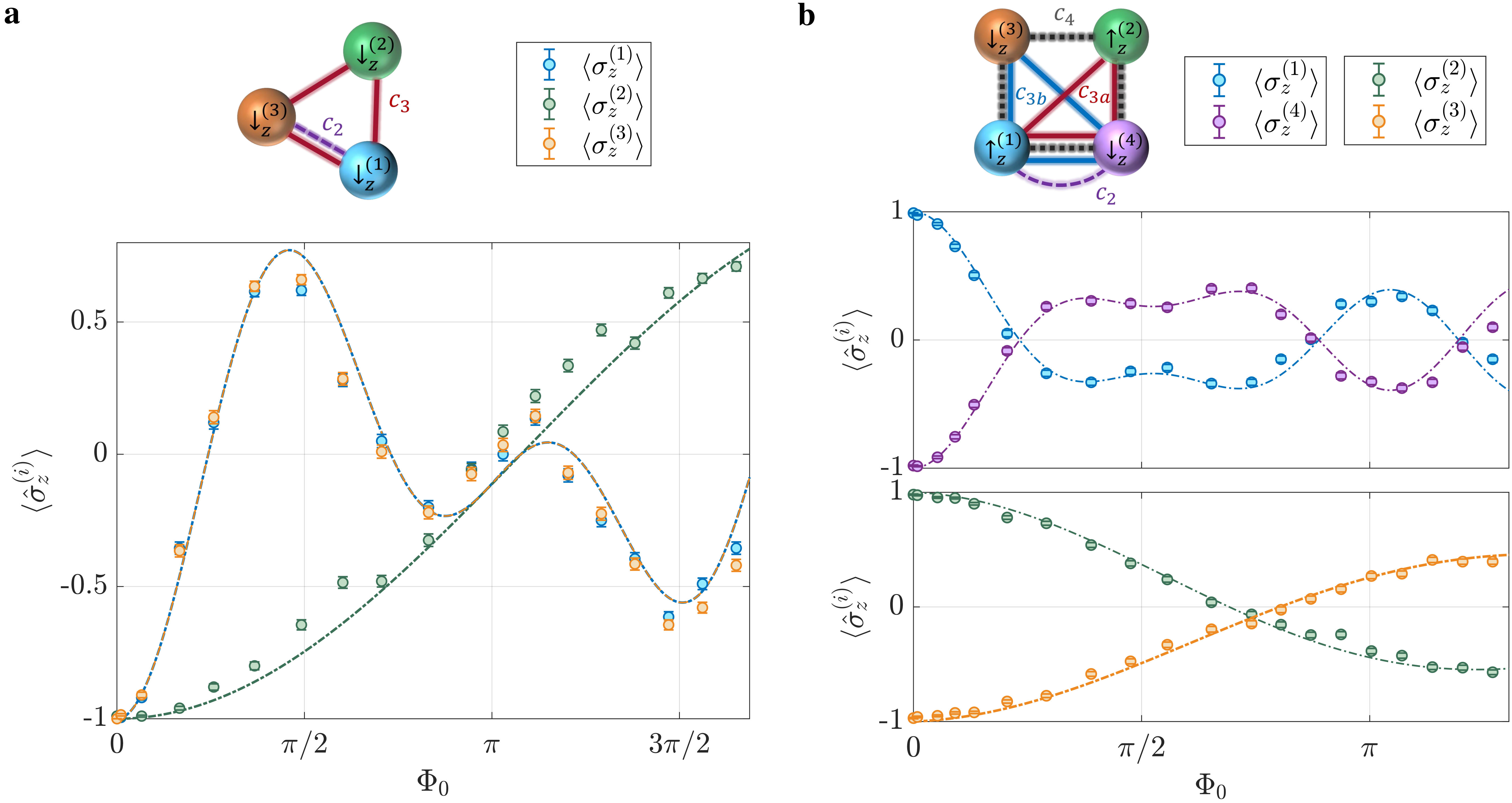}
\par\end{centering}
\centering{}\caption{\textbf{Effective Hamiltonians with three- and four-body interactions.} \textbf{a}, Evolution of three spins by the effective Hamiltonian $H_{\textrm{eff}}=\hbar\Phi/T$ for $\Phi$ in Eq.~(\ref{eq:Phi_3spins}) for the initial state $\ket{\downarrow_z^{(1)}\downarrow_z^{(2)}\downarrow_z^{(3)}}$. The effective Hamiltonian comprises two- and three-body terms whose magnitudes $c_2=1.03$ and $c_3=0.23$ as illustrated via the links connecting the different spins. The evolution as a function of $\Phi_0$ is realized with a fixed sequence time. \textbf{b}, Evolution of four spins by the effective Hamiltonian in Eq.~(\ref{eq:Phi_4spins}) containing simultaneously two-, three- and four-body terms with amplitudes $c_2=1.1$, $c_{3\textrm{a}}=0.36$, $c_{3\textrm{b}}=0.31$ and $c_4=0.1$. Dot-dashed lines are the analytically calculated magnetizations for the same initial state $\ket{\uparrow_z^{(1)}\uparrow_z^{(2)}\downarrow_z^{(3)}\downarrow_z^{(4)}}$ where the $c$ amplitudes are determined from the calculate squeezing parameters. Bars represent $1\sigma$ binomial uncertainties. The applied experimental sequences are presented in \ref{fig:sequences}b-c.
\label{fig:eff_Hamiltonian_evolution}}
\end{figure*}

We perform a limited characterization of this pure three-body interaction by measuring the output state distribution given each of the eight distinct three-qubit input eigenstates, all in the $z$ basis. The ideal population distributions of the expected GHZ-type states are equal weightings of the two complementary three-qubit states for each input state, as shown as wire frames in Fig.~\ref{fig:gate_tomo}a. The measured spin population distributions are shown as solid bars (see Supplementary Fig.~3 for the numerical values), resulting in an average population fidelity of $(95.8\pm0.9)\%$, uncorrected for SPAM and single-qubit gate errors. We further study the coherence in this three-body mapping from two particular input states $\ket{\downarrow_z^{(1)}\downarrow_z^{(2)}\downarrow_z^{(3)}}$ and $\ket{\uparrow_z^{(1)}\uparrow_z^{(2)}\uparrow_z^{(3)}}$ into the expected GHZ states $\tfrac{1}{\sqrt{2}}(\ket{\downarrow_z^{(1)}\downarrow_z^{(2)}\downarrow_z^{(3)}}\pm\ket{\uparrow_z^{(1)}\uparrow_z^{(2)}\uparrow_z^{(3)}})$. We measure the entanglement of these particular GHZ states using the parity fringe witness observable \cite{sackett2000experimental} as shown in Fig.~\ref{fig:gate_tomo}b, and extract state fidelities $\mathcal{F}=(94.8\pm1.5)\%$ and $\mathcal{F}=(94.4\pm1.9)\%$ respectively, uncorrected for SPAM and single-qubit gate errors. We estimate that the leading sources of errors are technical, and are noise in the beams' amplitude, motional noise of the oscillator, and uncompensated Stark-shifts.

Our scheme allows the realization of a continuous set of gates produced by the unitary evolution under an effective Hamiltonian $H_{\textrm{eff}}=\hbar\Phi/T$ at an effective time $T$ that is independent of the actual gate time, but instead scales linearly with $\Phi_0$. We demonstrate the evolution by the effective Hamiltonian associated with Eq.~(\ref{eq:Phi_3spins}) for $\xi=0.23$ (calculated from Eq.~\ref{eq:xi}) and for the initial state $\ket{\downarrow_z^{(1)}\downarrow_z^{(2)}\downarrow_z^{(3)}}$, presenting the magnetization $\langle\sigma_{z}^{(n)}\rangle$ of each spin in Fig.~\ref{fig:eff_Hamiltonian_evolution}. The observed spin evolution manifests interference effects owing to the interplay between the two- and three-body terms in the Hamiltonian, and is in good agreement with the analytically calculated evolution (dot-dashed lines). 

We extend this technique to generate an effective Hamiltonian in a four-ion chain. As in the three-ion gate, we act on two edge ions with displacement operations, while the squeezing is performed simultaneously on the two middle ions, with squeezing parameters $\xi$ and $\zeta$ as shown in \ref{fig:sequences}c. Unlike the displacement operation that is linear in the spin operators, the total scaling factor of the phonons, $e^{\xi\sigma_x^{(2)}}e^{\zeta\sigma_x^{(3)}}$, is multiplicative in the spin operators, owing to the nonlinear nature of the squeezing operation. Therefore, we realize an effective Hamiltonian that is associated with the geometric phase of the scaled rectangle:
\begin{equation}
\begin{aligned}\label{eq:Phi_4spins}\Phi=\Phi_0\Bigl(&c_2\sigma_{x}^{(1)}\sigma_{x}^{(4)}+c_{3\textrm{a}}\sigma_{x}^{(1)}\sigma_{x}^{(2)}\sigma_{x}^{(4)}\\+&c_{3\textrm{b}}\sigma_{x}^{(1)}\sigma_{x}^{(3)}\sigma_{x}^{(4)}+c_{4}\sigma_{x}^{(1)}\sigma_{x}^{(2)}\sigma_{x}^{(3)}\sigma_{x}^{(4)}\Bigr),\end{aligned}\end{equation} 
with two-, three- and four-body terms having relative amplitudes $c_2=\cosh\xi\cosh\zeta$, $c_{3\textrm{a}}=\sinh\xi\cosh\zeta$, $c_{3\textrm{b}}=\cosh\xi\sinh\zeta$ and $c_{4}=\sinh\xi\sinh\zeta$. In Fig.~\ref{fig:eff_Hamiltonian_evolution}b we demonstrate the evolution by this effective Hamiltonian for the initial state $\ket{\uparrow_z^{(1)}\uparrow_z^{(2)}\downarrow_z^{(3)}\downarrow_z^{(4)}}$ and the applied values $\xi=0.34$ and $\zeta=0.29$ (as calculated from Eq.~\ref{eq:xi}), following calibration of $\Phi_0$. The evolution in this case manifests interference between the four different terms in the Hamiltonian, and is in good agreement with the theoretically calculated evolution (dot-dashed lines).

In summary, we demonstrate a technique to realize native entangling operations comprised of higher order interactions between the spins of trapped ions. Our approach allows engineering new classes of programmable native gates and Hamiltonians using current trapped-ion hardware and requiring only minor alternations in the optical force spectrum and modest levels of squeezing.

It is interesting to compare this approach with the alternative necessary resources of a digital quantum computer using just two-qubit gates(and neglecting the cost of single qubit operations). The approach presented here allows the preparation of effective Hamiltonians comprising families of polynomials of Pauli strings in the $x$ basis whose order is up to the length $N$ of the chain. Owing to the collective nature of the phonon modes that are used as a quantum bus, these polynomials can feature dense connectivity, resulting in many different and nonlocal interaction terms. This scheme requires a fixed amount of displacement operations (equivalent to several two qubit gates) and additionally squeezing operations that carry a run-time overhead whose relative duration for a fixed optical power grows linearly in $N$ \cite{Katz2022Nbody}. Given only up to two-qubit operations, construction of a single Pauli string of order $n>2$ requires $2n$ two-qubit gates \footnote{see section 4.7.3 in \cite{Mike_and_Ike}}. A general spin polynomial with commuting terms requires a sequential application of $\sum_{n=2}^{N}{\tbinom{N}{n}2n}$ two-qubit gates, which grows exponentially in $N$ \cite{bullock2004asymptotically}. While some polynomials can be efficiently constructed with two qubit gates, the dramatically different scaling suggests that the operations  presented here can potentially speedup many operations in a quantum processor.

This demonstration can be extended to a variety of different sequences which can improve the robustness of the operations and/or construct  different sets of spin entangling operations. For example, the rectangular-shaped loop can be alternatively shaped to other trajectories in phase-space that would render the operation more robust to noise (e.g.~to frequency drifts of the oscillator) using pulse-shaping techniques \cite{shapira2018robust,wang2020high,leung2018robust,webb2018resilient,shapira2022robust}. Furthermore, pulse-shaping of the squeezing pulses enables control over the coupling between spins and all motional modes, despite the density of the second-sideband spectrum, enabling extension of the technique to longer ion chains \cite{katz2022programmable}.  

While the displacement and squeezing operations in this demonstration were executed sequentially and on resonance, simultaneous \cite{Katz2022Nbody} and/or off-resonance operations \cite{katz2022programmable} can natively realize additional other classes of quantum gates and spin Hamiltonians comprising high order interactions.
These many-body interactions are generally accompanied by other many-body terms of lower order, as demonstrated above (Fig.~\ref{fig:eff_Hamiltonian_evolution}b). The full power and expression of such interactions in trapped ion quantum computers may thus benefit from machine learning approaches \cite{Biamonte2017} to deploying families of such interactions to speedup and improve the performance of general quantum circuits.

\setcounter{figure}{0} 
\renewcommand{\figurename}{}
\renewcommand{\thefigure}{Extended Data Fig.~\arabic{figure}}
\begin{figure*}[t]
\begin{centering}
\includegraphics[width=14cm]{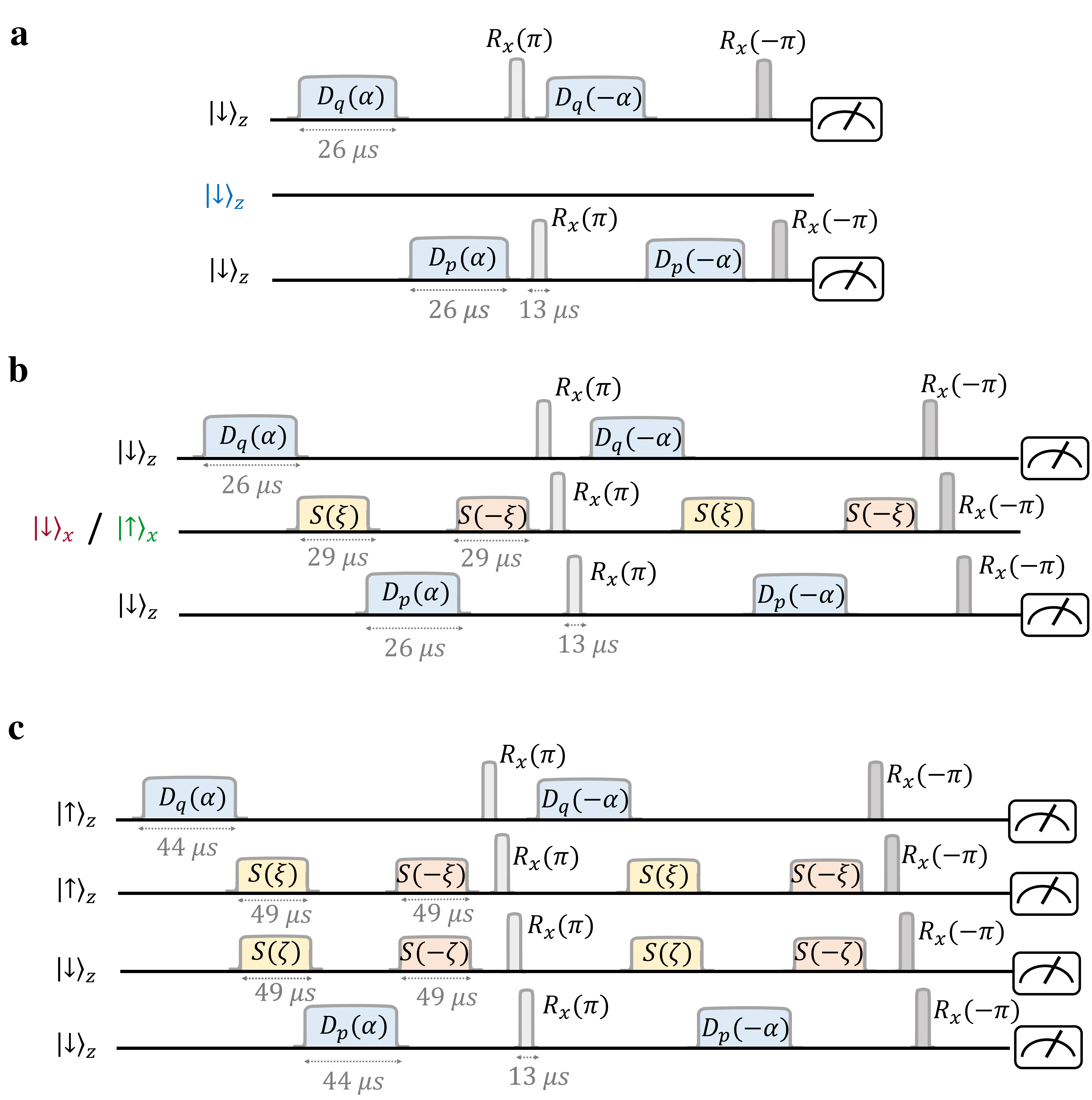}
\par\end{centering}
\centering{}\caption{\textbf{Experimental sequences.} \textbf{a}, Sequence of displacement operations acting on the two edge ions and composing the MS interaction, enclosing a closed rectangular loop in phase-space and generating the evolution Fig.~\ref{fig:phase_space_area}a. \textbf{b}, Superimposing spin-dependent squeezing operations on the second spin scales the displacement generated by the third spin by a factor $\exp(\sigma_x^{(2)}\xi)$ and consequently also the enclosed phase-space area. This sequence was applied to the configurations in Fig.~\ref{fig:phase_space_area}b-c and in Fig.~\ref{fig:gate_tomo} and Fig.~\ref{fig:eff_Hamiltonian_evolution}a. \textbf{c}, Displacement of the edge two spins and simultaneous squeezing of the middle spins for the four ions configuration presented in Fig.~\ref{fig:eff_Hamiltonian_evolution}b. The simultaneous squeezing scales the displacement generated by the fourth spin by a spin-dependent factor $\exp(\sigma_x^{(2)}\xi+\sigma_x^{(3)}\zeta)$ as seen from the identity $S^{(m)}(-\xi)D_p^{(n)}(\pm\alpha)S^{(m)}(\xi)=D_p^{(n)}(\pm e^{\sigma_x^{(m)}\xi}\alpha)$. The operators $D_p^{(n)}(\pm\alpha)$ and $D_q^{(n)}(\pm\alpha)$ denote displacement of the target phonon mode via the $n$th ion by $\pm\alpha$ along the $p$ and $q$ coordinates respectively. $S^{(m)}(\pm\xi)$ denotes the squeezing operator acting on ion $m$ and $R^{(n)}_{\theta}(\pm\pi)$ denotes short single-qubit $\pi$-pulses acting on the $n$ th ion, which commute with the spin-dependent displacement operations and which correct for slowly-varying uncompensated Stark shifts without altering the target state. See Methods for further details. \label{fig:sequences}}
\end{figure*}

\begin{acknowledgments}
This work is supported by the ARO through the IARPA
LogiQ program; the NSF STAQ and QLCI programs; the DOE QSA
program; the AFOSR MURIs on Dissipation Engineering
in Open Quantum Systems, Quantum Measurement/
Verification, and Quantum Interactive Protocols; the
ARO MURI on Modular Quantum Circuits; and the DOE HEP QuantISED Program.
\end{acknowledgments}

\clearpage
\part*{\centerline{Methods}}

\section*{Native ion-phonon interactions}\label{sec:appendix_A}
We control the displacement and squeezing operations of the ions using pairs of optical Raman beams. The beam that globally addresses the chain traverses an Acousto-Optical Modulator (AOM) that is simultaneously driven with two RF signals $A_+\sin(\omega_\textrm{+}t+\phi_+(t))$ and $A_-\sin(\omega_{-}t+\phi_-(t))$ that split and shift its optical frequency into two distinct tones. We simultaneously drive both red and blue sideband transitions to generate displacement or squeezing operations, and set their amplitudes $A_+$ and $A_-$ to be nearly equal. Control over the beat note frequency $\tfrac{1}{2}(\omega_{+}-\omega_{-})$ of the tones with respect to the carrier transition enables selection of the driven sidebands transitions; here we address the lowest frequency radial mode, denoted as mode number $1$ with frequency $\omega_1$, by tuning the relative detuning $\Delta=\tfrac{1}{2}(\omega_{+}-\omega_{-})-\omega_1$ to be on resonance ($\Delta=0$). 
We modulate the amplitude of the individually addressing beams to control the amplitude of the displacement of the target zig-zag mode and also to suppress the displacements of all other off-resonant modes. The amplitude-shaped waveform is based on the optimal-control technique \cite{katz2022programmable} using a sinusoidal basis and is exemplified in Supplementary Figure 1 for a single displacement stage of a three ion chain. 

The spin-dependent displacement operator on the $n$-th ion as a function of time is given by \cite{Katz2022Nbody}\begin{equation} \label{eq:displacement_operator}D^{(n)}(\alpha(t))=\exp\left(\sigma_{\varphi}^{(n)}\bigl(\alpha(t) \hat{a}^\dagger  -\alpha^*(t) \hat{a}\bigr)\right),\end{equation} 
where $\hat{a}=\hat{q}+i\hat{p}$ denotes the bosonic phonon annihilation operator of the target lowest-frequency radial mode, and the time dependent displacement is given by \begin{equation} \label{eq:alpha_disp} \alpha(t)=\tfrac{1}{2}\eta_{nk}\int_0^{t}d\tau e^{i(\Delta t+\delta{\varphi})}\Omega_n(\tau). \end{equation} 
Here $\Omega_n(t)$ is the Rabi frequency that can be varied up to about $1$ MHz, as we independently calibrate experimentally using single qubit rotation acting on the carrier transition. The Lamb-Dicke parameters $\eta_{nk}=0.08b_{nk}$ characterizing the spin-motion coupling, include mode participation factors for the zigzag mode ($k=1$, indexed in an ascending frequency order) of $b_{n1}=\left(0.41,0.82,0.41\right)$ for the three ion chain and $b_{n1}=\left(0.21,-0.67,0.67,-0.21\right)$ for the four ion chain, assuming a quadratic trapping potential along the chain axis. The applied potentials give an average spacing of about $3.7\,\mu$m between the ions to maximize the coupling with the equidistant fixed-spacing optical Raman beams, for which $\omega_1=2.817$ MHZ for the three ion chain and $\omega_1=2.781$ MHz for the four ion chain, where the radial center-of-mass frequency in both configurations is $3.03$ MHz. 

The relative phase between the two RF tones $\delta\varphi=(\phi_{+}-\phi_{-})/2$ controls the orientation of displacement in phase-space, and the common phase $\bar{\varphi}=(\phi_{+}+\phi_{-}-\pi)/2$  determines the orientation of the spin operator on the Bloch sphere $\sigma_{\bar{\varphi}}^{(n)}=\cos\bar{\varphi}_n\sigma_{x}^{(n)}+\sin\bar{\varphi}_n\sigma_{y}^{(n)}$. We nominally tune the phase $\bar{\varphi}_n=0$ to render the operator $\sigma_{\bar{\varphi}}^{(n)}=\sigma_{x}^{(n)}$ as considered in the main text, and tune $\Delta\equiv\Delta_1=0$ to resonantly drive the lowest frequency phonon-mode and generate the edges of the rectangular-shape trajectories whose orientation depends on the relative phase; motion along the $q$ ($-q$) coordinate (denoted as $D_q$ in \ref{fig:sequences}) is realized by setting the relative phase at $\delta\varphi=0$ ($\delta\varphi=\pi$) and motion along $p$ ($-p$) coordinate (denoted as $D_p$ in \ref{fig:sequences}) is realized by setting $\delta\varphi=\pi/2$ ($\delta\varphi=3\pi/2$). The duration of a displacement pulse is $26\,\mu$s in the three ions configuration and $44\,\mu$s in the four ions configuration.

We apply squeezing operations on the zig-zag mode by tuning the relative frequency of the tones at twice the motional frequency $\Delta=\tfrac{1}{2}(\omega_\textrm{+}-\omega_\textrm{-})-2\omega_1$=0. 
Here we use an unmodulated square-shaped pulse (with about ~$1\,\mu$s for the rise-time and fall-time of the edges), and we find that the expected coupling to other modes is small. This acts on the $n$-th ion to squeeze the motion using according to the operator,
\begin{equation}
S_{\hat{\xi}}^{(n)}(t)=e^{\frac{1}{2}\hat{\sigma}_{\bar{\phi}}^{(n)}\xi(t)(\hat{a}^{2}-\hat{a}^{\dagger2})}, 
\end{equation}
where the projection of the spin operator is controlled by the common phase of the two tones, $\bar{\phi}=\pi+(\phi_{+}+\phi_{-})/2$. We nominally tune $\bar{\phi}=0$ so that $\sigma_{\bar{\phi}}^{(n)}=\sigma_{x}^{(n)}$. The squeezing amplitude and its orientation is thus given by the complex parameter
\begin{equation}\label{eq:xi}
\xi(t)=\tfrac{1}{2}\eta_{n1}^2\int_0^td\tau\Omega_n(\tau)e^{i(\Delta\tau+\delta\phi)}.
\end{equation}
We tune $\Delta=0$ to resonantly drive the zig-zag mode and control the direction in the motional phase-space that is squeezed using $\delta\phi=(\phi_{+}-\phi_{-})/2$. The direction of squeezing in the experiment is aligned with the $q$ coordinate, with squeezing (anti-squeezing) corresponding to $\delta\phi=0$ $(\pi)$. The duration of a squeezing pulse is about $29\,\mu$s in the three ions configuration and about $49\,\mu$s in the four-ion configuration.

We implement single qubit rotations using a composite pulse sequence (SK1 sequence, see \cite{brown2004arbitrarily}) to implement the unitary operation \begin{equation}R^{(n)}_\theta(\chi)=\exp\left(-i\frac{\chi}{2}\sigma_{\theta}^{(n)}\right)\end{equation} on the $n$-th spin by driving a single tone of the global beam ($A_-=0$) on resonance with the carrier transition such that $\theta=\phi_+$. Each single qubit gate takes $12.7\,\mu$s. We use these single-qubit operations for preparation of the initial state along the $x$ axis and for echo pulses $R^{(n)}_0(\pm\pi)$ that commute with the spin operators in the circuit while suppressing effects of small uncompensated light shifts.  

\section*{Experimental Calibration}\label{sec:appendix_A}
We verify the value of $\bar{\varphi}=0$ by running a short sequence $R^{(n)}_\theta(\tfrac{\pi}{2})D^{(n)}(\alpha)R^{(n)}_\theta(-\tfrac{\pi}{2})$ on the target spin initially in state $\ket{\downarrow_z}$ and for variable rotation axis $\theta$. The angle $\theta_0$ for which the spin-flip probability is minimal, gives $\bar{\varphi}=\theta_0+\tfrac{\pi}{2}$. Similarly, we verify the value of $\bar{\phi}=0$ by running a short sequence $R^{(n)}_\theta(\tfrac{\pi}{2})S^{(n)}_{\xi}R^{(n)}_\theta(-\tfrac{\pi}{2})$ on the target spin initially $\ket{\downarrow_z}$, and find that $\bar{\varphi}=\theta_0$.

We verify the relative orientation of the displacement and squeezing operations, $\delta\varphi$ with respect to $\delta\phi$, by measuring the spin flip probability of the edge ions for the squeezed rectangle phase-gate by the sequence in \ref{fig:sequences}b for the initial state $\ket{\downarrow_z^{(1)}\downarrow_x^{(2)}\downarrow_z^{(3)}}$ and the nominal value of $\delta\varphi=0$ but for different values of $\delta\phi$. For small displacement and squeezing ($\Phi_0e^{\xi}<\tfrac{\pi}{2}$), the spin-flip probability is minimized for $\delta\phi=0$ and is given by $p_a=\sin^2(e^{-\xi}\Phi_0)$, whereas it is maximized for $\delta\phi=\pi$ and is given  $p_b=\sin^2(e^{\xi}\Phi_0)$. The measured values of $p_a$ and $p_b$ determine $\Phi_0$ and $\xi$.

We compensate for light-induced shifts generated during the displacement and squeezing operations by tuning the relative amplitude imbalance of the two tones  $|A_+-A_-|/|A_++A_-|$ at the level of about $1\%$. The imbalance applied during the squeezing and displacement operations is tuned independently to account for different shifts, e.g.~from residual coupling to Zeeman states outside of the space of the spin qubits and detuned $\pm4.2$ MHz from the carrier spin flip transition. The orientation of the magnetic field is also tuned to suppress the Zeeman transition amplitudes.

We routinely calibrate for drifts in the motional frequency of the oscillator by driving sequences that combine $D^{(n)}(\alpha)D^{(n)}(-\alpha)$ and scanning for the motional frequency $\Delta$ as shown in Supplementary Figure 2a. Here we implement $D^{(n)}(-\alpha)$ by applying the pulse $D^{(n)}(\alpha)$ while shifting the motional phase by $\delta\varphi\rightarrow\delta\varphi+\pi$, which manifests the inverse of the displacement operation only if the driving is resonant ($\Delta=0$). When the displacement is driven off-resonantly, the operation is not reversed and residual coupling between the spin and phonons leads to nonzero spin flip probability, which allows calibration of the motional frequency to better than $100$ Hz. Similarly, the operations $S^{(n)}_{\xi}S^{(n)}_{-\xi}$ where $S^{(n)}_{-\xi}$ are realized by advancing $\delta\phi$ by $\pi$, giving a similar dependence on the motional frequency. We generated a composite sequence that uses these operations and allows for efficient check of the motional frequency as shown in Supplementary Figure 2b.    

\bibliography{Refs}

\end{document}